\documentclass[12pt]{amsart}
\usepackage{amsmath}

\newtheorem{thm}{Theorem}
\newtheorem{lem}[thm]{Lemma}
\newtheorem{prop}[thm]{Proposition}
\newtheorem{cor}[thm]{Corollary}

\theoremstyle{definition}

\theoremstyle{remark}
\newtheorem{remark}{Remark}[section] 

\theoremstyle{plain}

\newcommand{\Z}{{\mathbf Z}}
\newcommand{\Q}{{\mathbf Q}}
\newcommand{\R}{{\mathbf R}}
\newcommand{\C}{{\mathbf C}}

\newcommand{\tr}{\operatorname{tr}}

\newcommand{\HH}{\mathbf H}

\newcommand{\TT}{\mathbf T^2}
\newcommand{\Ind}{\operatorname{Ind}}

\newcommand{\OPN}{\operatorname{Op}_N}
\newcommand{\OPh}{\operatorname{Op}_h}
\newcommand{\OP}{\operatorname{Op}}

\newcommand{\Hh}{\mathcal H_h}
\newcommand{\HN}{\mathcal H_N}
\newcommand{\TN}{T_N}  
\newcommand{\Tp}{T^{(p)}}       

\newcommand{\Uh}{U_h}
\newcommand{\UN}{U_N}
\newcommand{\Up}{U^{(p)}}  
\newcommand{\Uev}{U^{(2)}} 

\newcommand{\Fh}{\mathcal F_h} 

\newcommand{\toh}{\xrightarrow[h\to 0]{}} 
\newcommand{\Mat}{\operatorname{Mat}} 

\newcommand{\D}{D}
\newcommand{\T}{\TN}
\newcommand{\PP}{P}

\newcommand{\OF}{{\mathfrak O}}
\newcommand{\OO}{\OF_K}
\newcommand{\OK}{\OO}
\newcommand{\vv}{\nu}
\newcommand{\nn}{n}
\newcommand{\w}{\UN}
\newcommand{\trace}{\tr}
\newcommand{\mnorm}{ {\mathcal N}}

\newcommand{\Torus}{{\mathcal C}_A}
\newcommand{\Torusev}{\Torus^{\theta}}
\newcommand{\TorusK}{{\mathcal C}_K}

\newcommand{\unit}{\alpha}
\newcommand{\rr}{r}

\numberwithin{equation}{section}

\begin{document}

\title [Hecke theory and equidistribution for quantized cat maps]
{Hecke theory and equidistribution for the quantization of linear maps
of the torus}
\author{P\"ar Kurlberg and Ze\'ev Rudnick}
\address{Raymond and Beverly Sackler School of Mathematical Sciences,
Tel Aviv University, Tel Aviv 69978, Israel.  
Current address: Department of Mathematics, 
University of Georgia, Athens, GA 30602, U.S.A.
({\tt kurlberg@math.uga.edu})}
\address{Raymond and Beverly Sackler School of Mathematical Sciences,
Tel Aviv University, Tel Aviv 69978, Israel 
({\tt rudnick@math.tau.ac.il})}

\date{August 5, 1999}
\thanks{Supported in part by grants from the Israel Science
  Foundation and the US-Israel Binational Science Foundation. 
  In addition, the first author was partially supported by
  the EC TMR network "Algebraic Lie Representations", EC-contract no
  ERB FMRX-CT97-0100}

\begin{abstract}

We study semi-classical limits of eigenfunctions of a quantized
linear hyperbolic automorphism of the torus (``cat map'').  
For some values of Planck's constant, 
the spectrum of the quantized map has large degeneracies. 
Our first goal in this paper is to show that these degeneracies are
coupled to the existence of {\em quantum symmetries}. There is a
commutative group of unitary operators  on the state-space which commute with
the quantized map and therefore act on its eigenspaces. 
We call these ``Hecke operators'', in analogy with the setting of
the modular surface. 

We call the eigenstates of both the quantized map  and of all the Hecke
operators  ``Hecke eigenfunctions''.  
Our second goal is to study the semiclassical limit of
the Hecke eigenfunctions. We will show that they become
equidistributed with respect to Liouville measure, that is the
expectation values of quantum observables in these eigenstates
converge to the classical phase-space average of the observable. 

\end{abstract}

\maketitle

\section{Introduction} 

\subsection{Background} 
One of the key issues of ``Quantum Chaos'' is the nature of 
the semi-classical limit of eigenstates of classically chaotic systems.
When the classical system is given by the
geodesic flow on a compact Riemannian manifold $M$  
(or rather, on its co-tangent bundle), one can formulate the problem 
as follows: 
The quantum Hamiltonian is, in suitable units, represented by 
the positive Laplacian $-\Delta$ on $M$. 
To measure the distribution of its eigenstates, 
one starts with a (smooth) classical observable, 
that is a (smooth) function on the unit 
co-tangent bundle $S^*M$, and   
via some choice of quantization from symbols to pseudo-differential
operators, forms its quantization $\OP(f)$. 
This  a zero-order pseudo-differential operator
with principal  symbol $f$. 
The expectation value of $\OP(f)$ in the eigenstate 
$\psi$ is $\langle \OP(f) \psi,\psi\rangle$. 

Let  $\psi_j$ be a sequence of normalized 
eigenfunctions: $\Delta \psi_j +\lambda_j \psi_j=0$, 
$\int_M |\psi_j|^2 = 1$. 
The problem then is to understand the possible limits as $\lambda_j\to
\infty$ of the distributions   
\begin{equation}\label{dphih}
f\in C^\infty(S^* M) \mapsto 
\langle \OP(f) \psi_j,\psi_j \rangle .
\end{equation}
In the case that the geodesic flow is ``chaotic'', it is assumed 
that the eigenfunctions are ``random'', for instance in the sense that 
the expectation values 
converge as $\lambda_j\to \infty$ to
the average of $f$ with respect to Liouville measure on $S^*M$. 
The validity of this for {\em almost all} eigenmodes if the classical flow is 
{\em ergodic} (so a very weak notion of chaos!) 
is asserted by ``Schnirelman's theorem'' \cite{Sch1} 
\footnote{see Zelditch \cite{Zelditch87} and Colin de Verdiere \cite{CdV} for
proofs.},  
a fact sometimes referred to as {\em quantum ergodicity}.  
The case where there are no exceptional subsequences 
is called {\em Quantum Unique Ergodicity} (QUE). Its validity 
seems to be a very difficult problem, to-date unsolved in any case
where the dynamics are truly chaotic 
(see however Marklof and Rudnick \cite{Marklof-Rudnick} 
where QUE is proved for an ergodic, though non-mixing, model case).


\subsection{Cat maps}
In order to shed some light on the validity of QUE, we look at a 
``toy model'' of the situation - the quantization of linear hyperbolic
automorphisms of the 2-dimensional torus $\TT$. Here the phase space
$\TT$ is {\em compact} and instead of a Hamiltonian flow we consider a
discrete time dynamics, generated by the iterations of a single map
$A\in SL(2,\Z)$. If $A$ is {\em hyperbolic}, that is $|\tr A|>2$, then
this map is a paradigm of chaotic dynamics. Such maps are sometimes
called ``cat maps'' in the physics literature. 
A quantization of these
``cat maps'' was proposed by Hannay and Berry \cite{BH} and elaborated
on in \cite{Knabe, DE, DEGI, KLMR, Zel-cat}. 
We review this in some detail in 
Sections~ \ref{background}, \ref{dynamics}.  
In particular the admissible values of Planck's constant are inverse
integers $h=1/N$, and the Hilbert space of states $\HN\simeq
L^2(\Z/N\Z)$ 
of the quantum system is finite dimensional, of dimension $N=h^{-1}$.
To every classical observable $f\in C^\infty(\TT)$ one 
associates an operator $\OPN(f)$ on $\HN$, the corresponding
quantum observable. 
The quantization of the cat map is a unitary operator $\UN(A)$ on
$\HN$, the quantum propagator, unique up to a phase factor,
characterized by an exact\footnote{This {\em exact} version of
Egorov's theorem is  very special and is a 
consequence of the map being linear.} version 
of Egorov's theorem 
\begin{equation}\label{int egorov}
\UN(A)^{-1}\OPN(f) \UN(A) = \OPN(f\circ A), \qquad \forall
f\in C^\infty(\TT)
\end{equation}

The eigenvectors $\phi$ of the quantum propagator $\UN(A)$ are the
analogues of the eigenmodes 
of the Laplacian, and to study their concentration properties
one forms  the  distributions 
\begin{equation*}
f\mapsto \langle \OPN(f) \phi,\phi \rangle 
\end{equation*}
In particular we want to understand the quantum limits as $N\to
\infty$.  
An analogue of Schnirelman's theorem in this setting was proven in 
\cite{BdB, Zel-cat}.
One would like to know if QUE holds, that is if
the only quantum limit is the uniform measure on $\TT$.  

The spectrum of the quantum propagator $\UN(A)$ has degeneracies,
which renders the study of possible quantum limits difficult. 
 The degeneracies are systematic and are inversely related to the
order of $A\mod 2N$.  
Degli~Esposti, Graffi and Isola \cite{DEGI} showed that if 
instead of looking at all integer values of $N$, 
one restricts to the sparse\footnote{It is an open problem to
show that there are infinitely many  primes where the degeneracy is 
bounded. This is known
assuming the Generalized Riemann Hypothesis, which in fact guarantees
that a positive proportion of  the primes satisfy the assumption.}   
subsequence consisting of primes for which the degeneracies are
{\em bounded}, and  moreover split in the quadratic
extension of the rationals containing the eigenvalues of $A$, then  
the only limit is indeed the uniform measure.  

Our first goal in this paper is to show that the degeneracies are
coupled to the existence of {\em quantum symmetries}. There is a
commutative group of unitary operators  on $\HN$ which commute with
$\UN(A)$ and therefore act on each eigenspace of $\UN(A)$. 
We will call these ``Hecke operators'', in analogy with the setting of
the modular surface\footnote{A notable difference between our
setting and the modular surface is that there one expects few, if any,
degeneracies.}  \cite{RS,LS, Jakobson94}. 
We may thus consider eigenfunctions of the desymmetrized quantum map, 
that is eigenstates of both $\UN(A)$ and of all the Hecke
operators. We call these {\em Hecke eigenfunctions}.  
Our second goal is to show that these become
equidistributed with respect to Liouville measure, that is the
expectation values of quantum observables in Hecke eigenstates
converge to the classical phase-space average of the observable.

\subsection{Results} We turn to a detailed description of our results. 
We first carry out a systematic study of the quantum propagator. We
define $\UN(A)$ so that it only depends on the remainder of 
$A\mod 2N$ and satisfies \eqref{int egorov}. 
One gets a {\em projective} representation $A\mapsto \UN(A)$ of
the subgroup of ``quantizable'' elements in the finite modular group 
$SL(2,\Z/2N\Z)$. We
explain (Section~\ref{multiplicativity}) 
that it can be made into an {\em ordinary} representation 
if we further restrict to the subgroup  $\Gamma(4,2N)$ given by 
$g=I \mod 4$ for $N$ even, $g=I \mod 2$ for $N$ odd. Thus for $A,B\in
\Gamma(4,2N)$ we have $\UN(AB)=\UN(A)\UN(B)$. Consequently, if
$AB=BA\mod 2N$ then their propagators commute.
This is the basic principle that we use to form the Hecke operators. 

Fix a hyperbolic matrix $A$, which we will further assume
lies in the congruence subgroup 
$$\Gamma(4)=\{ g\in SL(2,\Z): g=I \mod 4\}$$
so that its reduction modulo $2N$ lies in
$\Gamma(4,2N)$ for all $N$. To find matrices  commuting with $A$
modulo $2N$, we use the connection with the theory of real quadratic
fields (Section~\ref{sec:Hecke ops}): 
If $\unit$ is an eigenvalue of $A$, form 
$\OF=\Z[\unit]$ which is an {\em order} in the real quadratic field
$K=\Q(\unit)$. There is an $\OF$-ideal $I$ so that the action of
$\unit$ on $I$ by multiplication has $A$ as its matrix in a suitable
basis. Thus the action of $\OF$ on $I$ by multiplication 
gives us an embedding 
$\iota:\OF\hookrightarrow \Mat_2(\Z)$, and induces a map $\iota:\OF/2N\OF\to
\Mat_2(\Z/2N\Z)$. Under this map, the images of elements  $\beta\in \OF/2N\OF$
whose Galois 
norm is $1\mod 2N$ lie in $SL(2,\Z/2N\Z)$ and commute with $A$ modulo
$2N$. If we further require that $\beta=1\mod 4\OF$ then we get
a group of commuting matrices  $\iota(\beta)\in\Gamma(4,2N)$, whose
quantum propagators  $\UN(\iota(\beta))$ commute with $\UN(A)$ and
with each other. These are our Hecke operators.  

Since the Hecke operators commute with $\UN(A)$, they act on its 
eigenspaces, and since they commute with each other there
is a basis of $\HN$ consisting of joint eigenfunctions of $\UN(A)$ and
the Hecke operators, whose  elements we call Hecke eigenfunctions. 
Our main theorem is
\begin{thm}\label{main thm of intro}
Let $A\in \Gamma(4)$ be a hyperbolic matrix, 
and $f\in C^\infty(\TT)$ a smooth observable. 
Then for all  normalized Hecke eigenfunctions $\phi\in \HN$ of
$\UN(A)$,  the expectation values 
$\langle \OPN(f)\phi,\phi \rangle$ converge to the phase-space
average of $f$ as $N\to \infty$. Moreover,
for all $\epsilon>0$ we have  
$$
\langle \OPN(f)\phi,\phi \rangle = \int_{\TT} f(x)dx + 
O_{f,\epsilon}( N^{-1/4+\epsilon}),\quad \text{as } N\to \infty
$$
\end{thm}
\begin{remark}
It is easy to extend Theorem~\ref{main thm of intro} 
to give similar results for matrix elements of $\OPN(f)$. When $N$ is
such that the degeneracies in the spectrum of $\UN(A)$ are sufficiently
small, this implies as in \cite{DEGI} that the expectation values of
$\OPN(f)$ in  {\em all} eigenstates converge to $\int_{\TT} f(x)dx$. 
\end{remark}
\begin{remark} 
The exponent of $1/4$ in our theorem is certainly not optimal, and
more likely the correct exponent is $1/2$. That is the exponent given
in \cite{DEGI}, where the problem is reduced to one-variable
exponential sums, 
which can be estimated using  Weil's theorem - the Riemann Hypothesis
for a curve over a finite field. 

What we in fact show
(Theorem~\ref{main thm for onb's}) is
that if $\phi_i$, $i=1,\dots,N$  is an orthonormal basis of
$\HN$ consisting of Hecke eigenfunctions then 
$$
\sum_{i=1}^N  \left\lvert\langle \OPN(f)\phi_i,\phi_i \rangle - 
\int_{\TT} f(x)dx \right\rvert^4  \ll N^{-1+\epsilon}
$$
from which we deduce Theorem~\ref{main thm of intro} by taking an
orthonormal basis with $\phi_1=\phi$ and 
omitting all but one term on the LHS. If all terms on the LHS are of
roughly the same size then we would expect this to give the exponent
$1/2$. 
\end{remark}

The proof of Theorem~\ref{main thm of intro} is reduced to a counting
problem in Section~\ref{ergodicity}. 
This in turn comes down to counting solutions of
the congruence
$$
\beta_1-\beta_2+\beta_3-\beta_4 = 0 \mod N\OF 
$$
in norm-one elements $\beta_i\in \OF/N\OF$. 
The number of such
norm-one elements is $O(N^{1+\epsilon})$ 
(Lemma~\ref{number of Hecke ops}), and  
since this equation has 3 degrees of freedom, the trivial bound of the
number of solutions is $O(N^{3+\epsilon})$, $\forall \epsilon>0$. 
To get any result in Theorem~\ref{main thm of intro} we need
to show that the number of solutions is $O(N^{3-\delta})$ for some
$\delta>0$, that is any saving over the trivial bound would do.  
This is accomplished in Section~\ref{sec: counting} 
where we show that the number of solutions is $O(N^{2+\epsilon})$, the
optimal bound.

\noindent{\bf Acknowledgements:} We thank J. Bernstein, 
D. Kazhdan, J. Keating, J. Marklof, F. Mezzadri, P. Sarnak and S. Zelditch 
for helpful discussions concerning various points in the paper.   
\newpage
\section{Background on quantization of maps}\label{background}

In this paper we consider the quantization of linear (orientation
preserving) 
automorphisms of the torus $\TT=\R^2/\Z^2$, that is 
elements of the modular group $SL(2,\Z)$, which for the most part
will be assumed hyperbolic (known as ``cat maps'' in some of the
literature). 
For this we first review a procedure (one of several) for
quantization of maps.  

The first to quantize the cat map were Hannay and Berry 
\cite{BH}. We will follow in part an approach by means of  
representation theory which was 
developed by Knabe {\cite{Knabe} and Degli~Esposti, Graffi and 
Isola \cite{DE, DEGI}. See also 
\cite{KLMR, BdB, Zel-cat} for other approaches.

\subsection{The quantization procedure} 
We start by describing some desiderata for a quantization procedure
for a symplectic map $A$ of a phase space.  
In the literature it is customary to distinguish two components of
the quantization procedure  -  a kinematic component and a dynamical
one.   

In the {\em kinematic} component one constructs a Hilbert space 
$\Hh$ of
states of the quantum system\footnote{$h$ stands for Planck's
constant.} and an algebra of operators on the space
- the algebra of quantum observables. 
Smooth functions  $f$ on the classical phase space of the
system (that is classical observables) 
are mapped to members
$\OPh(f)$ of this algebra. To make the connection with the
classical system, 
it is required that in the  limit $h\to 0$, the
commutator of the quantization of two observables $f$, $g$ reproduce the
quantization of their Poisson bracket 
$\{f,g\} = \sum_j \frac{\partial f}{\partial p_j}
\frac{\partial g}{\partial q_j} -\frac{\partial f}{\partial q_j}
\frac{\partial g}{\partial p_j} $: 
\begin{equation}\label{poisson}
\frac i{\hbar} \left [\OPh(f),\OPh(g)\right ]-\OPh(\{f,g\}) \toh 0
\end{equation}
(we won't specify the sense of convergence).

The {\em dynamical} part of quantization amounts to prescribing a
discrete time evolution of the algebra of quantum observables, 
that is 
a unitary map $\Uh(A)$ of $\Hh$, 
which reproduces  the classical map $A$ in the
limit $h\to 0$ in the sense that: 
\begin{equation}\label{egorov}
\Uh(A)^{-1} \OPh(f) \Uh(A) -\OPh(f\circ A) \toh 0
\end{equation}
(this is the analogue of Egorov's theorem). 

In our case, the classical phase space is the torus $\TT$. The classical
observables are smooth functions on $\TT$. 
We will find that Planck's
constant $h$ is restricted to be an inverse integer: $h=1/N$, $N\geq 1$. 
The state-space $\Hh$ will be $\HN=L^2(\Z/N\Z)$. To each observable 
$f\in C^\infty(\TT)$ we will assign, by an analogue of Weyl quantization,  
an operator $\OPN(f)$ on 
$\HN$ so that \eqref{poisson} holds, where convergence is in
the space of $N\times N$ matrices. 
The dynamics will be given by a linear map $A\in SL(2,\Z)$ so that 
$x=(\begin{smallmatrix}p\\q\end{smallmatrix})\in
\TT\mapsto Ax$ is a symplectic map of the torus. 
Given an observable $f\in C^\infty(\TT)$, the
classical evolution defined by $A$ is $f\mapsto f\circ A$, where 
$f\circ A(x)=f(Ax)$. It turns out that for a certain subset of
matrices $A$,  there is a unitary map $\UN(A)$ on $L^2(\Z/N\Z)$
so that an exact form of \eqref{egorov} holds: 
$$ 
\UN(A)^{-1}\OPN(f)  \UN(A) =\OPN(f\circ A) , \qquad \forall
f\in C^\infty(\TT)
$$
This will be our discrete time evolution.

Below we describe these procedures in detail. 

\subsection{Kinematics: The space of states} \label{kinematics} 
As the Hilbert space of states, we take distributions $\psi(q)$ on the
line $\R$ which are periodic in both the position and the momentum
representation. As is well known, this restricts Planck's constant to
take only inverse integer values. 
We review the  argument: Recall that the momentum
representation of a wave-function $\psi$ is  
$$
\Fh \psi(p) = 
\frac 1{\sqrt{h}}\int_{-\infty}^\infty \psi(q)e^{-2\pi i\frac{qp}{h}} dq
$$ 
We then  require
$$
\psi(q+1)=\psi(q), \qquad \Fh\psi(p+1)= \Fh \psi(p)
$$
(one may just require that this hold up to a phase). 
 From periodicity  in the position representation, we get 
$$
\psi(q) = \sum_{n\in \Z} c_n e(nq)
$$
where 
$$e(z):=e^{2\pi i z}$$
In the momentum representation, that is applying $\Fh$, we get 
$$
\Fh\psi(p) = \sqrt{h}\sum_{n\in \Z} c_n  \delta( p -nh)
$$
Now in order that $\Fh\psi(p+1)= \Fh\psi(p)$ we clearly need  
$\frac 1h \in \Z$, that is for some integer $N\geq 1$ that 
$$
h=\frac 1N
$$
In that case we also need 
$$c_{n+N} = c_n$$

Thus one finds that $h=1/N$ and the space of states is finite
dimensional, of dimension $N=1/h$, and consists of periodic 
point-masses at the coordinates  $q=Q/N$, $Q\in \Z$. 
We may then identify $\HN$ with the 
$N$-dimensional vector space $L^2(\Z/N\Z)$, with
the inner product $\langle\,\cdot\,,\,\cdot\,\rangle$ defined by
\begin{equation*}
\langle \phi,\psi \rangle 
 = \frac1N \sum_{Q\bmod N} \phi(Q) \, \overline\psi(Q) ,
\end{equation*}

\subsection{Quantizing observables} 
Next we construct  quantum observables: 
For a free particle on the line, we would take as the basic observables 
the position and momentum operators 
$$
\hat q \psi(q):=q\psi(q),\qquad 
\hat p\psi(q):=\frac \hbar i \frac{d\psi}{dq}(q)
$$ 
($\hbar = h/2\pi$). 
 For our periodic phase space we take the basic observables to be 
$e(\hat q)=e^{2\pi i\hat q}$ and  $e(\hat p)$, 
which correspond to the phase space translations 
$$e(\hat q)\psi(q) = e(q)\psi(q),\qquad e(\hat p)\psi(q) =\psi(q+h)$$
Corresponding to the commutation relation 
$$
[\hat q,\hat p] =  i\hbar  = - \frac h{2\pi i}
$$
we find that 
$$
e(\hat q) e(\hat p) = e^{-2\pi i h}  e(\hat p) e(\hat q) 
$$
Writing 
$$ t_1 :=e(\hat p), \qquad t_2:= e(\hat q) $$
(so that $t_2t_1=e^{-2\pi i h}t_1t_2$) we put for $n=(n_1,n_2)\in \Z^2$
\begin{equation}\label{phase space translation}
\TN(n) := e^{\frac {i\pi n_1 n_2}N} t_2^{n_2} t_1^{n_1}
\end{equation}
Their action on a wave-function $\psi\in L^2(\Z/N\Z)$ is 
\begin{equation}\label{action of T(n)}
\TN(n)\psi(Q) = e^{\frac {i\pi n_1 n_2}N} e(\frac{n_2Q}N)\psi(Q+n_1)
\end{equation}
These are clearly of period $2N$ in $n$:
$$
\TN(n+2Nm)=\TN(n),\quad n,m \in \Z^2
$$
The adjoint of $\TN(n)$ is given by 
\begin{equation}\label{adjoint T(n)}
\TN(n)^* = \TN(-n)
\end{equation}
They also satisfy 
\begin{equation}\label{T(m+n)}
\TN(m)\TN(n) = e^{\frac {i\pi \omega(m,n)}N} \TN(m+n)
\end{equation}\
where $$\omega(m,n) = m_1n_2-m_2 n_1$$

Now we can finally construct quantum observables: 
For any smooth classical observable 
$f\in C^\infty(\TT)$ with  Fourier expansion 
$$f(x) = \sum_{n\in \Z^2} f_n e(n\cdot x),\quad 
x=(\begin{smallmatrix}p\\q\end{smallmatrix})\in \TT,
$$  
we define its quantization $\OPN(f)$ as
$$
\OPN(f) := \sum_{n\in \Z^2} f_n \TN(n)
$$
The verification of \eqref{poisson} is  an easy calculation using
\eqref{T(m+n)}. 

\subsection{The Heisenberg group}\label{heisenbergsec}
We now digress to connect this construction to the representation
theory of a certain Heisenberg group $H_{2N}$. 

For vectors $x=(x_1,x_2)$, $y=(y_1,y_2)$ define   
$\omega(x,y):=x_1y_2-x_2y_1$. This is a non-degenerate symplectic form. 
The Heisenberg group $H_{2N}$ is defined to be the set
$(\Z/2N\Z)^2\times\Z/2N\Z$ with multiplication  
$$
(x,z)\cdot (x',z'):=(x+x',z+z'+\omega(x,x'))
$$
This is at odds with the standard convention where one multiplies 
$\omega$ by $1/2$, but is essential for us because $2$ is not
invertible in $\Z/2N\Z$. 

It is useful to record various facts about the multiplication in
$H_{2N}$:  
The inverse of $(x,z)$ is 
\begin{equation} 
(x,z)^{-1} = (-x,-z) 
\end{equation}
The commutator of two elements is given by 
\begin{equation} 
(x,z)(x',z')(x,z)^{-1}(x',z')^{-1} = (0,2\omega(x,x'))
\end{equation}
 From this commutator identity and the fact that $\omega$ is
non-degenerate we immediately find
\begin{lem}
The center of $H_{2N}$ is $(N\Z/2N\Z)^2\times \Z/2N\Z$, that is 
$$
Cent(H_{2N}) = \{(N\epsilon,N\eta,z):\epsilon,\eta=0,1,
\quad z\in \Z/2N\Z \}
$$
\end{lem}

We define a representation of $H_{2N}$ on  $L^2(\Z/N\Z)$ by setting  
$$
\pi(n,z) = e(\frac z{2N})\TN(n)
$$
 From the relation \eqref{T(m+n)} it follows that
$\pi(h)\pi(h')=\pi(hh')$, i.e. we do indeed get a representation. 

The center of $H_{2N}$ then acts via the character $\chi$ given by  
$$
\chi(x_0,y_0,z) = e(\frac{z+x_0y_0}{2N})
$$
(that is $\pi(x_0,y_0,z) = \chi(x_0,y_0,z) I$). 

The basic facts about $\pi$ and the representation theory of $H_{2N}$ 
are  
\begin{prop}
i) All irreducible representations of $H_{2N}$  have dimension at most
$N$. 

ii) The representation $\pi$ is irreducible, and is the unique irreducible
$N$-dimensional representation with central character $\chi$. 
\end{prop}

We omit the details of the proof; the main point (which is easy to
verify from the definitions) is 
\begin{lem}\label{l:trace} 
the trace of $\TN(n)$ is given by 
$$
|\trace \T(n) | = 
\begin{cases}
  N & \text{if $n \equiv (0,0) \mod N$,} \\
  0 & \text{otherwise.}
\end{cases}
$$
\end{lem}
\begin{proof}
Let $\phi_i = \sqrt{N} \delta_i$ where $\delta_i$ is the Dirac delta function
supported at $i$, so that $\{ \phi_i\}_{i=1}^N$ is an
orthonormal basis of $L^2(\Z/N\Z)$. Then
$$
\tr \TN(n) = \sum_{i=1}^{N} \langle\TN(n) \phi_i,  \phi_i \rangle
$$
and by equation~\eqref{action of T(n)}
\begin{equation*}
\begin{split}
 \TN(n) \phi_i  (Q) &=
e( \frac{ n_1 n_2 + 2 n_2 Q}{2N} ) \phi_i(Q + n_1) \\
&=
e( \frac{ n_1 n_2 + 2 n_2 Q}{2N} ) \phi_{i-n_1}(Q) \\
&=e( \frac{- n_1 n_2 + 2 n_2 i}{2N} ) \phi_{i-n_1}(Q)
\end{split}
\end{equation*}
Therefore $\tr\TN(n)=0$ unless 
$n_1 \equiv 0 \mod N$, in which case 
$$
\sum_{i=1}^{N} \langle\TN(n) \phi_i,  \phi_i  \rangle 
= 
e( \frac{- n_1 n_2 }{2N} ) \sum_{i=1}^{N} e( \frac{ n_2 i}{N} ).
$$
The result now follows since $\sum_{i=1}^{N} e( \frac{ n_2 i}{N} )$
equals $N$ if $n_2 \equiv 0 \mod N$, and is zero otherwise. 
\end{proof}


\subsection{Description of $\pi$ as an induced representation} 
Let $Y$ be the subgroup of elements 
$$Y=\{(x_0,y,z):y,z\in \Z/2N\Z,x_0\in N\Z/2N\Z\}$$ 
It is easily seen to be a normal, maximal abelian subgroup, 
of index $N$, containing the center. Set for $(x_0,y,z)\in Y$ 
$$
\tau(x_0,y,z):=e(\frac {z+x_0 y}{2N})
$$
This is a character of $Y$ (we need to use $2x_0\equiv 0\mod 2N$ in
verifying this), restricting to the character
$\chi(x_0,y_0,z)= e(\frac {z+x_0 y_0}{2N})$ of the center. 

We consider the induced representation $\Ind_Y^{H_{2N}} \tau$  
of the Heisenberg group. The basic model for it is the space of
functions $\Phi:H_{2N}\to \C$ satisfying 
$\Phi(a h)=\tau(a)\Phi(h)$ for $a\in Y$, $h\in
H_{2N}$. The action of the group is by right multiplication
$h\Phi(h'):=\Phi(h'h)$. 
By restricting to the subgroup $X=\{(x,0,0)\}$ we can 
realize this induced representation as functions on $\Z/2N\Z$ which
are $N$-periodic (since the element $(N,0,0)$ lies in $X\cap Y$). 
This space of functions we can identify with $L^2(\Z/N\Z)$. 

Let us compute the action of a group element $h=(x,y,z)\in H_{2N}$ in
this model. For this we need to write $(x',0,0)\cdot h$ as 
$a\cdot (x'',0,0)$, $a\in Y$. The relevant identity is  
$$
(x',0,0)(x,y,z) = (0,y ,z+ xy+ 2 x' y)(x'+x,0,0)
$$
Thus the element $h=(x,y,z)$ acts as 
$$
h\phi(x') = e(\frac{z+xy+2x'y}{2N})\phi(x'+x)
$$
In particular $(x,0,0)$ acts as translation by $x$ and $(0,y,0)$ as 
a multiplication operator 
$\phi(x')\mapsto e(\frac{x'y}{N})\phi(x')$. 
The center acts by the character $(x_0,y_0,z)\mapsto
e(\frac{z+x_0y_0}{2N})$. 
These show  that $\pi$ coincides with the induced representation
$\Ind_Y^{H_{2N}} \tau$.

\newpage

\section{Dynamics: quantized cat maps}\label{dynamics}
We now show how to assign to (certain)  linear automorphisms
$A$ of the torus $\TT$ a unitary operator $\UN(A)$ on
$L^2(\Z/N\Z)$, which satisfies:  For all observables $f\in
C^\infty(\TT)$
$$ 
 \UN(A)^{-1}  \OPN(f) \UN(A)= \OPN(f\circ A),\qquad
$$

The finite modular group $SL(2,\Z/2N\Z)$ acts by automorphisms on the
Heisenberg groups $H_{2N}$ via $(x,z)^A:=(xA,z)$,  $A\in SL(2,\Z/2N\Z)$. 
That this is indeed an automorphism 
(that is $(h_1h_2)^A=h_1^A h_2^A$) 
follows from $A$ preserving the symplectic form $\omega$. Moreover we
have $(h^A)^B = h^{AB}$. Composing the representation $\pi$ of
$H_{2N}$ with $A$ gives 
a new representation  $\pi^A(h):=\pi(h^A)$, which is clearly still an
irreducible $N$-dimensional representation. Its central character
$\chi^A$ can be easily computed as follows: if $x_0,y_0\in N\Z/2N\Z$
and $(x_1,y_1)=(x_0,y_0)A$ then $\chi^A$ is given by 
$$
\chi^A(x_0,y_0,z) = \chi((x_0,y_0)A,z) = e(\frac{z+x_1y_1}{2N})
$$
This will be the same character as $\chi$ iff $x_1y_1\equiv x_0y_0
\mod 2N$ for all $x_0,y_0\in N\Z/2N\Z$. Writing 
$A=(\begin{smallmatrix} a&b\\c&d \end{smallmatrix})$ 
and
$x_0 = N\epsilon$, $y_0=N\eta$, $\epsilon,\eta\in \Z/2\Z$, this is
equivalent to requiring 
$$
N(ab\epsilon^2+cd \eta^2) \equiv 0\mod 2,\qquad 
\forall \epsilon,\eta\in \Z/2\Z
$$
or 
$$
Nab\equiv Ncd \equiv 0\mod 2
$$
This is only a restriction if $N$ is odd, and is  satisfied by
the elements of the theta group 
$$
\Gamma_\theta(2N)=\{\begin{pmatrix} a&b\\c&d \end{pmatrix}\in SL(2,\Z/2N\Z):
ab\equiv cd\equiv 0 \mod 2\}
$$

Therefore if $A\in \Gamma_\theta(2N)$, we get a unitarily equivalent
representation $\pi^A$ of $H_{2N}$. Thus there is a unitary map
$\UN(A)$, the {\em quantum propagator} associated to $A$, so that 
$$
\pi(h^A) = \UN(A)^{-1}\pi(h)\UN(A), \qquad 
\forall h\in H_{2N}
$$
In particular we find 
\begin{equation}\label{exact egorov for T(n)}
\UN(A)^{-1}  \TN(n) \UN(A) =  \TN(n A)
\end{equation}
and consequently for all observables $f\in C^\infty(\TT)$, 
\begin{equation}\label{exact egorov for OP(f)}
\OPN(f\circ A) = \UN(A)^{-1}  \OPN(f) \UN(A) 
\end{equation}

We now for any ``quantizable'' element $A\in SL(2,\Z)$ (that is
$A=(\begin{smallmatrix} a&b\\c&d \end{smallmatrix})$ with $ab\equiv
cd\equiv 0\mod 2$), we define the {\em quantum propagator} 
(or ``quantized cat map'')  to be $\UN(\bar A)$ where $\bar A\in
SL(2,\Z/2N\Z)$ is the reduction of $A$ modulo 
$2N$. Thus {\em by its construction}, $\UN(A)$ only depends on the
reduction $A \mod 2N$. (This is a difference from the construction in 
Hannay and Berry \cite{BH}).

\newpage

\section{Multiplicativity}\label{multiplicativity}

The quantum propagators $\UN(A)$ are uniquely defined up to a phase-factor,
because of the irreducibility of $\pi$ (Schur's lemma). 
Thus they define  a {\em projective} representation of $\Gamma_\theta(2N)$,
that is 
$$\UN(AB)=e^{i\phi_N(A,B)}\UN(A)\UN(B)\qquad A,B \in
\Gamma_\theta(2N) 
$$
Define the subgroup 
$$
\Gamma(4,2N) = 
\left\{ g\in SL(2,\Z/2N\Z): \begin{cases} g=I \bmod 4,& N \text{ even}  \\
g=I \bmod 2,& N \text{ odd} \end{cases} \right\}
$$

The goal of this section is to show that there is a choice of phases
for the propagators $\UN(A)$ so that on the subgroup $\Gamma(4,2N)$ 
the map $A\mapsto \UN(A)$ is a homomorphism:
\begin{thm}\label{thm splitting}
There is a choice of quantum propagators 
so that   
$$
\UN(AB) = \UN(A)\UN(B),\quad A,B\in \Gamma(4,2N)
$$
\end{thm}
As a consequence we find 
\begin{cor}\label{cor commute}
If $A,B\in \Gamma(4,2N)$ commute mod $2N$
then their propagators also commute: $\UN(A)\UN(B) =
\UN(B)\UN(A)$. 
\end{cor}

Theorem~\ref{thm splitting} in various guises is essentially known,
and arose out of the study of theta-functions and the Weil
representation. One form is due to Kubota \cite{Kubota} (see also
\cite{Gelbart}). There are also treatments purely at the finite level
\cite{Nobs, Balian et Itzykson}. Since Corollary~\ref{cor commute} is
absolutely crucial to our work, and we did not find a good reference
for the exact form that we need, we will sketch a  proof
(or more precisely, a verification) of Theorem~\ref{thm splitting}. 
We wish to note that Theorem~\ref{thm splitting} is 
{\em a-priori} more subtle  than Corollary~\ref{cor commute}, since
once we know that there is {\em some} choice of phases for which
Corollary~\ref{cor commute} holds, than it holds for {\em all} choices;
this is not the case with Theorem~\ref{thm splitting}.\footnote{We
thank Jon Keating for emphasizing this point to us.}

\subsection{Reduction to prime powers} 
Factor $2N =\prod_{p}p^{k_p}= 2^{k}\prod_{p>2} p^{k_p}=2^{k}M$, 
$M$ odd. 
The Chinese remainder theorem gives an isomorphism 
$$
\Z/2N\Z \simeq \prod_{p} \Z/p^{k_p}\Z
$$ 
given by 
$$
x\mapsto (x\bmod p^{k_p})_p
$$
with inverse 
$$
(x_p \bmod p^{k_p})_p \mapsto \sum \frac{2N}{p^{k_p}} r_p x_p \bmod 2N
$$
where $r_p$ is the inverse of $2N/p^{k_p}$ modulo $p^{k_p}$. 
Correspondingly we have a bijection
$$
L^2(\Z/2N\Z) \simeq \bigotimes_p L^2( \Z/p^{k_p}\Z)
$$
We define the phase space translations $\Tp$ on $L^2(\Z/p^{k_p}\Z)$ as
in \eqref{action of T(n)},  by
$$
\Tp(n)\psi(Q) = e(\frac{r_p(n_1n_2+2n_2 Q)}{p^{k_p}})\psi(Q+n_1)
$$
Is is then a simple matter to see that $\TN(n)=\otimes_p \Tp(n)$, that
is if $\psi=\otimes_p \psi_p \in  \bigotimes_p L^2( \Z/p^{k_p}\Z)$ is
decomposable then 
$$
\TN(n)\psi(Q) = \prod_p \Tp(n)\psi(Q\bmod p^{k_p})
$$
This allows us to express the quantum propagators $\UN(A)$ as tensor
products. Indeed, if we already 
have propagators $\Up(A)$ which satisfy 
\begin{equation}\label{egorovp}
\Up(A)^{-1} \Tp(n) \Up(A) = \Tp(nA)
\end{equation}
We then set 
\begin{equation}\label{factorize U}
\UN(A) := \otimes \Up(A)
\end{equation}
which still  satisfies
$$
\UN(A)^{-1} \TN(n) \UN(A) = \TN(nA)
$$
for all $n\in \Z^2$ and therefore $\UN(A)$ coincides up to a phase
with any other map satisfying this.

We use this procedure to define $\UN(A)$ (that is, choose a phase) so
that $U_N$ is an honest representation of a  subgroup $\Gamma(4,2N)$
of $SL(2,\Z/2N\Z)$, not merely a projective representation. 
 From the factorization property \eqref{factorize U}, it follows that
it enough to show that $\Up$ is a representation of $SL(2,\Z/p^{k_p}\Z)$
when $p>2$ is odd, and   of $\Gamma(4,2^{k})$ if $N=2^{k-1} M$ is even. 

\subsection{Gauss sums} 
We need some preliminary information on Gauss sums. We define
normalized Gauss sums  
\begin{equation}\label{gaussum}
S_r(a,p^k) = 
\frac 1{\sqrt{p^k}} \sum_{x\bmod p^k} e(\frac {-ra x^2}{p^k})
\end{equation}
For $p$ odd these are 4-th roots of unity. To describe them, 
define for $t\in (\Z/p^k\Z)^*$ 
$$
\Lambda_{r,p^k}(t) = \frac{S_r(t,p^k)}{S_r(1,p^k)}
$$
Note that if $t=t_1^2\in (\Z/p^k\Z)^*$ is a square then
$\Lambda_{r,p^k}(t)=1$ since from \eqref{gaussum} we find after
the change of variables $x_1=t_1 x$ that $S_r(t,p^k)=S_r(1,p^k)$.

For odd $p$, $\Lambda_{r,p^k}$ is given in terms of the Legendre symbol as 
$$
\Lambda_{r,p^k}(t)  = \left(\frac tp\right)^k
$$
and is a character of $(\Z/p^k\Z)^*$: 
$$
\Lambda_{r,p^k}(tt')=\Lambda_{r,p^k}(t)\Lambda_{r,p^k}(t')
$$
When $p=2$, one has 
$$
\Lambda_{r,2^k}(t) = \left(\frac {-2^k}{t} \right) 
i^{-r(\bar t^2-1)/8}
$$
where $\bar t$ is the smallest positive residue of $t\bmod 4$. 
In that case it is not quite a character of the whole multiplicative
group of $\Z/2^k\Z$, but instead satisfies
\begin{equation}\label{hilbertsymb}
\Lambda_{r,2^k}(tt')=(t,t')_2\Lambda_{r,2^k}(t)\Lambda_{r,2^k}(t')
\end{equation}
where $(t,t')_2$ is the Hilbert symbol. In particular, 
if $t,t'=1 \bmod 4$ then the Hilbert symbol is trivial and so we get a
character of the subgroup $\{t=1\bmod 4\}\subset (\Z/2^k\Z)^*$ 
\footnote{this is relevant for $k\geq 2$.} 
given simply by 
$$
 \Lambda_{r,2^k}(t) = \begin{cases} 1,& t=1\mod 8\\ (-1)^k,& t=5 \mod 8
                       \end{cases}  
$$

For $p$ odd we will also need to know the normalized Gauss
sum \eqref{gaussum} when $t=-1$ in which case one has 
$$
S_r(-1,p^k) = \begin{cases}1,& k \text{ even}\\
\epsilon(p)\left(\frac rp \right), &k \text{ odd}\end{cases}
$$
where 
$$
\epsilon(p) = \begin{cases}1,&p=1\bmod 4\\ i,&p=3\bmod 4 \end{cases}
$$

\subsection{$p$ odd} We describe how to define $\Up$ on
$SL(2,\Z/p^k\Z)$ so that it gives a representation - see Nobs \cite{Nobs}
for details. This group is generated by the matrices 
\begin{equation}\label{gensodd}
\begin{pmatrix} 1&b\\ &1 \end{pmatrix},\quad 
\begin{pmatrix} t&\\ &t^{-1} \end{pmatrix},\quad  
\begin{pmatrix} &1\\ -1& \end{pmatrix}
\end{equation}
and so it suffices to specify $\Up$ on such matrices, provided we
preserve all relations between them. 
This is done by the formulas 
\begin{equation}\label{U(1b01)}
\Up\begin{pmatrix} 1&b\\ &1 \end{pmatrix}\psi(x) =
e(\frac{rbx^2}{p^k})\psi(x)
\end{equation}
\begin{equation}\label{U(t001/t)}
\Up\begin{pmatrix} t&\\ &t^{-1} \end{pmatrix}\psi(x) =
\Lambda_{r,p^k}(t) \psi(tx)
\end{equation}
\begin{equation}\label{U(w)}
\Up\begin{pmatrix} &1\\ -1& \end{pmatrix}\psi(x) = 
S_r(-1,p^k) \frac 1{\sqrt{p^k}}
\sum_{y\bmod p^k}\psi(y)e(\frac {2rxy}{p^k})
\end{equation}
It is easy to check that these satisfy \eqref{egorovp}. 
To see a verification that this prescription does indeed give a
{\em consistent} definition (that is that all relations between the 
generators \eqref{gensodd} are satisfied), see e.g. \cite{Nobs}. 
Once we have this then automatically we get $\Up(AB) = \Up(A)\Up(B)$.

\begin{remark} 
It is in fact the case that {\em any} projective representation of
$SL(2,\Z/p^k\Z)$, $p$ odd, can be modified to give a
representation (and more generally, $SL(2,\Z/m\Z)$ if $ m\neq 0\bmod 4$)- this
is due to Schur \cite{Schur} when $k=1$. See 
\cite{Mennicke} and \cite{Beyl} for the general case.  
\end{remark}

\subsection{$\mathbf{p=2}$} 
Here we restrict to the subgroup $\Gamma(4,2^{k})$, $k\geq 2$. 
The literature in this case is harder to come by, so we include
complete proofs. We start by describing generators and relations for
this group. 
More generally, let $p$ be any prime, and $k\geq 2$.
Let 
$$
\Gamma(p^2,p^k): = \{g\in SL(2,\Z/p^k \Z): g= I \mod p^2 \}. 
$$

\begin{lem}  \label{presentation}
$\Gamma(p^2,p^k)$ has a presentation with generators 
$u_+(x)$, $u_-(y)$, 
$s(t)$, where $x,y,t\in \Z/p^k\Z$, $x,y\equiv 0 \bmod p^2$, 
$t\equiv 1 \bmod p^2$,
and relations   
\begin{eqnarray}
u_+(x)u_+(x') &=& u_+(x+x') \label{rel1} \\
u_-(y) u_-(y') &=& u_-(y+y')  \label{rel2} \\
s(t)s(t') &=& s(tt')  \label{rel3} \\
s(t)u_+(x)s(t)^{-1} &=& u_+(t^2x) \label{rel4} \\
s(t)u_-(y)s(t)^{-1} &=& u_-(t^{-2}y) \label{rel5} \\
s(d)u_+(a)u_-(b) &=& u_-(d^{-1}b)u_+(da), \qquad d:=(1+ab)^{-1} \label{rel*}
\end{eqnarray}
\end{lem}
\begin{proof} 
Let $G$ be the abstract group with the above presentation. 
We get a map $\Psi$ from $G$ into $\Gamma(p^2,p^k)$ by taking  
$$
\Psi: u_+(x)\mapsto \begin{pmatrix} 1&x\\ &1 \end{pmatrix},\quad 
u_-(y)\mapsto \begin{pmatrix} 1&\\ y&1 \end{pmatrix},\quad  
s(t)\mapsto \begin{pmatrix} t&\\ &t^{-1} \end{pmatrix}
$$
One verifies that the relations hold in $SL(2,\Z/p^k\Z)$ so that 
$\Psi$ is a homomorphism. Next, note that we have a ``Bruhat
decomposition'' for $\Gamma(p^2,p^k)$: Every element can be {\em uniquely}
written in the form 
$$
\gamma=\begin{pmatrix} t&\\ &t^{-1} \end{pmatrix} 
\begin{pmatrix}1&x\\ &1 \end{pmatrix} 
\begin{pmatrix} 1&\\ y&1 \end{pmatrix}
$$
which follows from the formula 
$$
\begin{pmatrix} a&b\\c&d  \end{pmatrix}  = 
\gamma=\begin{pmatrix} d^{-1}&\\ &d \end{pmatrix} 
\begin{pmatrix}1& bd\\ &1 \end{pmatrix}
\begin{pmatrix} 1&\\ \frac cd&1 \end{pmatrix}
$$
(note that since $d=1 \mod p^2$ it is in particular invertible). 
This implies that the map $\Psi$ is surjective. 
To see that $\Psi$ is an isomorphism, 
it suffices to show that every element of the abstract
group $G$ can also be written in the form $g=s(t)u_+(x)u_-(y)$, since
then by the {\em uniqueness} of the decomposition in $\Gamma(p^2,p^k)$,
$\Psi$ is also one-to-one.

With the aid of the first five relations, 
every word $W \in G$ can be written as a product:
$$
W= s(t_1) u_+(x_1)u_-(y_1) \cdot \dots \cdot s(t_n) u_+(x_n)u_-(y_n)
$$
for some $n\geq 1$. 
We prove by induction on $n$ that we can write $W=s(t)u_+(x)u_-(y)$
for $x,y=0 \mod p^2$, $t=1 \bmod p^2$. When $n=1$ this holds
trivially, and for $n>1$ we use the relation \eqref{rel5}, \eqref{rel*} 
to  write 
$$
u_-(y_{n-1})s(t_n)u_+(x_n) = s(t_n)u_-(t_n^2 y_{n-1})u_+(x_n) = 
s(t_n)s(t')u_+(x')u_-(y')
$$
and so
\begin{equation*}\begin{split}
W&= s(t_1) u_+(x_1)u_-(y_1) \dots  s(t_{n-1}) u_+(x_{n-1}) 
s(t_n)s(t')u_+(x')u_-(y')u_-(y_n) \\
&= s(t_1) u_+(x_1)u_-(y_1) \dots 
s(t'_{n-1})u_+(x''_{n-1})u_-(y''_{n-1})
\end{split}\end{equation*}
after a further application of the first  five relations. The result now
follows by induction. 
\end{proof}

We now specify the propagators $\Uev(A)$ for the generators:
For $\begin{pmatrix}1&a\\ &1 \end{pmatrix}$ and  
$\begin{pmatrix} t&\\ &t^{-1} \end{pmatrix}$ they are given by  the
same formulas \eqref{U(1b01)}, \eqref{U(t001/t)}. For the matrices 
$$
\begin{pmatrix} 1&\\ b&1 \end{pmatrix}=  
\begin{pmatrix} &1\\ -1& \end{pmatrix}^{-1} 
\begin{pmatrix}1&-b\\ &1 \end{pmatrix}
\begin{pmatrix} &1\\ -1& \end{pmatrix}
$$
we  conjugate \eqref{U(1b01)} by an analogue of the Fourier transform
\eqref{U(w)}  and define 
\begin{equation}\label{U(10b1)}
\Uev\begin{pmatrix} 1&\\ b&1 \end{pmatrix}\psi(x)=  
\sum_{y\bmod 2^k}\psi(y) \frac 1{2^{k}}\sum_{z\bmod 2^k} 
e(\frac{r(-bz^2+2z(y-x) ) }{2^k})
\end{equation}

To show that this defines a representation, one has to check that all
the relations of Lemma \ref{presentation} are satisfied. The first
five are fairly 
straight-forward, bearing in mind that $\Lambda$ is a character of the
multiplicative group of residues $t=1\bmod 4$ (see \eqref{hilbertsymb}).  
The last relation \eqref{rel*} requires verifying an identity of Gauss sums:  
Unwinding the action of the right and left
hand sides in \eqref{rel*} we must show that
\begin{multline*}
\Lambda(d) 
\sum_{ z \mod 2^{k}} \sum_{ y \mod 2^{k}}
\psi(y)
e \left( \frac{\rr}{2^{k}} \left( 
  2yz - bz^2 - 2dxz + a d^2 x^2
\right) \right) \\=
\sum_{ z \mod 2^{k}} \sum_{ y \mod 2^{k}}
\psi( y)
e \left( \frac{\rr}{2^{k}} \left( 
  2yz - d^{-1}bz^2 - 2xz + a d y^2
\right) \right)
\end{multline*}

Now, $d \equiv 1 \mod 16$ implies that $\Lambda(d)=1$ since then $d$ 
is a square modulo $2^k$, and if the
identity is to hold for all $\psi$ and all values of $x$ we obtain that
for all $x,y$

\begin{multline}\label{weird gauss sum identity}
\sum_{ z \mod 2^{k}} 
e \left( \frac{\rr}{2^{k}} \left( 
  - bz^2 +  2z(y-dx) + a d^2 x^2
\right) \right) \\
=
\sum_{ z \mod 2^{k}} 
e \left( \frac{\rr}{2^{k}} \left( 
  - d^{-1}bz^2 
  + 2z(y-x) + a d y^2
\right) \right).
\end{multline}
We will verify this in Appendix~\ref{appgauss}. 

\newpage
\section{Hecke operators}\label{sec:Hecke ops}

%
%
%
%
%
We now introduce a commutative group of unitary operators on
$L^2(\Z/N\Z)$  which commute with $\UN(A)$. 
For this, we have to bring in the theory of quadratic fields 
(see \cite{PV} for a survey in connection to cat maps).

\subsection{Integral matrices and quadratic fields}
\label{s:cft}
Let $A \in SL_2(\Z)$ be a hyperbolic matrix: $|\tr A|>2$. 
The eigenvalues $\unit,\unit^{-1}$ of $A$ generate a field extension 
$K= \Q(\unit)$, which is a real quadratic field since
$\trace(A)^2>4$. 
We denote by $\OK$ the ring of integers of $K$. 
The eigenvalues $\unit,\unit^{-1}$ of $A$ will be units in $\OK$. 
Adjoining $\unit$ to $\Z$ gives an {\em order} $\OF=\Z[\unit]\subseteq
\OK$ in $K$.  
We claim that there is an $\OF$-ideal $I\subset\OF$  so
that the action of $\unit$  by multiplication on $I$ is equivalent to
the action of $A$ on $\Z^2$, in the sense that there is  a basis of $I$ with
respect to which the matrix of $\unit$ is precisely $A$. 

The construction is as follows \cite{Taussky}: 
Since $\unit$ is an eigenvalue of
$A$, there is a vector $v=(v_1,v_2)$ such that 
$vA = \unit v$ and $v\in\OF^2$. 
Let  $I:=\Z[v_1,v_2] \subset \OF$. Then $I$  is in an $\OF$-ideal, and
the matrix of $\unit$ acting on $I$ by 
multiplication in the basis $v_1,v_2$ is precisely $A$. 
\begin{remark} 
It is easy to check that the above construction sets up a bijection between
$GL_2(\Z)$-conjugacy classes of elements in $SL_2(\Z)$ with
eigenvalues $\unit$, $\unit^{-1}$ and 
and ideal classes in the order $\OF$. 
(Recall that two ideals $I_1,I_2$ are said to
be in the same ideal class if there exist nonzero $a,b \in \OF$ so
that $a I_1 = b I_2$.)
%
\end{remark}

In the same way, the action of $\OF$ by multiplication on $I$ gives us 
an embedding  
$$
\iota:\OF\hookrightarrow \Mat_2(\Z)
$$ 
so that $\gamma = x+y\unit \in \OF$ corresponds to $xI+yA$. Moreover, the
determinant of $xI+yA$ equals $\mnorm(\gamma)=\gamma \bar\gamma$, 
where $\mnorm : K \rightarrow \Q$ is the Galois norm. 
In particular, if $\gamma \in \OF$
has norm one then $\gamma$ corresponds to an element in $SL_2(\Z)$,
and if in addition $\gamma \equiv 1 \mod 4 \OF$ then $\gamma$
corresponds to an element in $\Gamma(4)$.

\subsection{Hecke operators}

Given an integer $M\geq 1$, the
embedding $\iota: \OF \hookrightarrow \Mat_2(\Z)$ induces a map
$\iota_M:\OF/M\OF\to \Mat_2(\Z/M\Z)$  and 
the norm $\mnorm:K\to \Q$ gives a
well-defined map 
$$
\mnorm : \OF/M\OF \rightarrow \Z/M\Z . 
$$ 
We let $\Torus(M)$ be the group of norm one elements in
$\OF/M\OF$. 
$$ 
\Torus(M) = \ker \left[ \mnorm:(\OF/M\OF)^*\to(\Z/M\Z)^* \right]
$$
Similarly, replacing the order $\OF$ by the maximal order $\OK$ we set
$$ 
\TorusK(M) = \ker \left[ \mnorm:(\OK/M\OK)^*\to(\Z/M\Z)^* \right]
$$
to be the norm one elements in $\OK/M\OK$. 

If $M=2N$ is even, we set $\Torusev(M)$ to be the elements of
$\Torus(2N)$ that are congruent to one modulo
$4\OF$ (respectively $2\OF$) if $N$ is even (resp. odd). For $M$ odd 
we set $\Torusev(M)=\Torus(M)$. 

By construction, the image of $\Torusev(2N)$ in $\Mat_2(\Z/2N\Z)$ lies in 
$\Gamma(4,2N)$. Since $\unit$ commutes with all elements in $\Torusev(2N)$ we
see that $A$ commutes, modulo $2N$, with the elements in
$\iota(\Torusev(2N))$. 
Thus by Corollary~\ref{cor commute} the quantizations $\UN(\iota(\beta))$ 
of $\beta\in\Torusev(2N)$ commute with $\UN(A)$ and with each other. 
We will call these ``Hecke operators''.

We will need to know the number of Hecke operators:
\begin{lem}\label{number of Hecke ops}
The number of elements of $\Torusev(2N)$ satisfies
$$
N^{1-\epsilon} \ll |\Torusev(2N)| \ll N^{1+\epsilon},\quad \forall
\epsilon>0 . 
$$
\end{lem}
\begin{proof}
Since the reduction map $\OF\to \OF/4\OF$ has image of size $4^2$,  
$\Torusev(2N)$ has bounded index in $\Torus(2N)$. 
The inclusion $\OF\subset \OK$ induces a map  $\OF/M\OF \to \OK/M\OK$
which has kernel and co-kernel of size at most $[\OK:\OF]$,
independent of $M$. Therefore the induced map 
$\Torus(M)\to \TorusK(M)$ 
on norm-one elements also has
bounded kernel and co-kernel. 
Thus it suffices to prove the lemma in the case of the maximal order $\OK$. 
By the Chinese remainder theorem, it suffices to prove it  
in the case of prime powers, which is given in Appendix~\ref{app: count} by 
Lemma~\ref{l:number-of-norm-one}. 
\end{proof}

\subsection{Hecke eigen-functions}
The Hecke operators $\UN(\iota(\beta))$, $\beta\in \Torusev(2N)$,
commute with each other and with $\UN(A)$. Therefore the eigen-spaces
of the unitary map $\UN(A)$ break up into joint eigen-spaces of the 
Hecke operators. Such a joint eigen-function we call a {\em Hecke
eigen-function}.  
In other words, there
exist an orthonormal basis $\{ \phi_i \}$ of $L^2(\Z/N\Z)$ and
characters $\lambda_i$ of $\Torusev(2N)$ such that $\phi_i$ are
eigenfunctions of $\UN(A)$ and 
$$
\w(\iota(\beta)) \phi_i = \lambda_i(\beta) \phi_i,  \quad  
\forall \beta \in \Torusev(2N).
$$
We call such a basis of $L^2(\Z/N\Z)$ a {\em Hecke basis}.  

\newpage
\section{Ergodicity of Hecke eigenfunctions}\label{ergodicity}

In this section and the following we show that if 
$\phi \in L^2(\Z/N\Z)$ 
is a normalized Hecke eigenfunction then the expectation values 
$\langle \OPN(f)\phi,\phi \rangle$ converge to the classical
phase-space average 
$\int_{\TT} f$ for all smooth observables 
(Theorem~\ref{main thm of intro} of the Introduction). 
In fact, we show something  stronger:  
\begin{thm}\label{main thm for onb's}
Let $\phi_i\in L^2(\Z/N\Z)$, $i=1,\dots,N$  be any orthonormal basis of Hecke 
eigenfunctions  of $\UN(A)$. Then 
$$
\sum_{i=1}^N  \left\lvert\langle \OPN(f)\phi_i,\phi_i \rangle - 
\int_{\TT} f(x)dx \right\rvert^4  \ll_{f,\epsilon} N^{-1+\epsilon}
$$
\end{thm}
\subsection{Proof of Theorem~\ref{main thm for onb's}} 
To prove this theorem, it suffices (see below) to prove it for the basic
observables $f(x) = e(nx)$, $0\neq n\in \Z^2$, that is to show
\begin{thm}\label{main thm for onb's for T(N)}
Let $0\neq n\in \Z^2$, and 
let $\phi_i\in L^2(\Z/N\Z)$, $i=1,\dots,N$  be any orthonormal basis
of Hecke  eigenfunctions  of $\UN(A)$. Then 
$$
\sum_{i=1}^N |\langle \TN(n)\phi_i,\phi_i \rangle |^4 \ll_\epsilon |n|^{16}
N^{-1+\epsilon} ,\quad N\to \infty
$$
\end{thm}
The proof of Theorem~\ref{main thm for onb's} from 
Theorem~\ref{main thm for onb's for T(N)} is easy using the rapid
decay of the Fourier coefficients of $f$. Indeed, write
$f(x)=\sum_{n\in \Z^2}\widehat f(n) e(nx)$, so that 
$\OPN(f)=\sum_{n\in \Z^2}\widehat f(n) \TN(n)$. Therefore 
\begin{multline*}
\sum_{i=1}^N  \left\lvert\langle \OPN(f)\phi_i^N,\phi_i^N \rangle - 
\int_{\TT} f(x)dx \right\rvert^4  = 
\sum_{i=1}^N   \left\lvert \sum_{0\neq n\in \Z^2}
\widehat f(n)\langle \TN(n)\phi_i,\phi_i \rangle \right\rvert^4 \\
\leq \sum_{i=1}^N   \sum_{n_1,\dots,n_4\neq 0} 
\prod_{k=1}^4 |\widehat f(n_k)\langle \TN(n_k)\phi_i,\phi_i \rangle|
\end{multline*}
For notational convenience we write 
$$
t_i(n):=|\langle \TN(n)\phi_i,\phi_i \rangle|
$$ 
Now interchange the order of summation, and apply Cauchy-Schwartz
twice: For fixed $n_1,n_2,n_3,n_4$ 
\begin{multline*}
\sum_{i=1}^N t_i(n_1)t_i(n_2)t_i(n_3)t_i(n_4) \leq \\ 
\left(\sum_{i=1}^N (t_i(n_1)t_i(n_2))^2 \right)^{1/2}
\left(\sum_{i=1}^N (t_i(n_3)t_i(n_4))^2 \right)^{1/2} 
\leq \prod_{k=1}^4\left(\sum_{i=1}^N t_i(n_k)^4 \right)^{1/4}
\end{multline*}
Now use Theorem~\ref{main thm for onb's for T(N)}: For $n_k\neq 0$, 
$$
\left(\sum_{i=1}^N t_i(n_k)^4 \right)^{1/4} \ll 
|n_k|^4 N^{-1/4+\epsilon}
$$
and so we get 
$$
\sum_{i=1}^N t_i(n_1)t_i(n_2)t_i(n_3)t_i(n_4)\ll N^{-1+\epsilon'}
\prod_{k=1}^4 |n_k|^4
$$
Now sum over all possible $n_k\neq 0$ to find 
$$
\sum_{i=1}^N  \left\lvert\langle \OPN(f)\phi_i,\phi_i \rangle - 
\int_{\TT} f(x)dx \right\rvert^4  \ll
N^{-1+\epsilon} \left( \sum_{n\neq 0} \widehat f(n)|n|^4 \right)^4
$$
which proves Theorem~\ref{main thm for onb's}. \qed

\subsection{Reduction to a counting problem} 
We first reduce Theorem~\ref{main thm for onb's for T(N)} to a
counting problem. 
\begin{prop} \label{prop reduce to count}
Fix $0\neq n=\iota(\nu) \in \Z^2$, $\nu\in I$. Then for any 
Hecke basis of eigenfunctions $\phi_i$,  
\begin{multline*}
\sum_{i=1}^N |\langle \TN(n)\phi_i,\phi_i \rangle |^4 \leq \\
\frac {N}{|\Torusev(2N)|^4}\#\{ \beta_i\in \Torusev(2N):
\nu(\beta_1-\beta_2+\beta_3-\beta_4) =0 \mod N I \}
\end{multline*}
\end{prop}

In order to prove Proposition~\ref{prop reduce to count}, we define
for $n=\iota(\nu)$, $0\neq \nu\in I$   
$$
D = D(n) = \frac{1}{|\Torusev(2N)|} 
\sum_{\beta \in \Torusev(2N)}
\w(\iota(\beta))^{-1} \T(n) \w(\iota(\beta)).
$$
If $(t_{ij})$ is the matrix coefficients of $\T(n)$ expressed in the
eigenvector basis $\{ \phi_k \}$ so that $t_{ij} =
<\T(n)\phi_i,\phi_j>$, then we see that
$$
D_{ij} =
\frac{1}{|\Torusev(2N)|} 
\sum_{\beta \in \Torusev(2n)}\lambda_i(\beta)\overline {\lambda_j(\beta)}
t_{ij} 
$$ 
Since the sum of a nontrivial character over all elements in a group
vanishes we have
\begin{equation}
\label{e:d-chi}
D_{ij} =
\begin{cases}
  t_{ij} & \text{if $\lambda_i =  \lambda_j$,} \\
  0 & \text{otherwise.}
\end{cases}
\end{equation}
%

\begin{lem}
\label{l:cutting}
With $D$ defined as above we have
$$
\sum_{ \lambda_i =  \lambda_j }  
|t_{ij}|^4
\leq
\trace( (D^*D)^2).
$$
\end{lem}
\begin{proof}
  Let $D =  (d_{ij}) = (v_i)$ where the $v_i$'s are the column
  vectors of $D$. Examining the $(k,k)$-entry  of
  $(D^*D)^2$ we get
$$ ( (D^*D)^2 )_{kk} = 
\sum_i <v_i,v_k><v_k,v_i> = 
\sum_i |<v_i,v_k>|^2,
$$ 
and hence 
$$
\trace( (D^*D)^2 ) \geq \sum_k |<v_k,v_k>|^2 \geq \sum_{i,j}
|d_{ij}|^4 .
$$
The result now follows from equation~\eqref{e:d-chi}.
\end{proof}

\begin{lem}
\label{l:trace-bound}
We have 
$$
\trace( (\D^{*}\D)^2 ) \leq 
\frac{N}{|\Torusev(2N)|^4}
|\{ \beta_i \in \Torusev(2N):
\nu(\beta_1-\beta_2+\beta_3-\beta_4) \equiv 0 \mod N I \}|
$$
\end{lem}
\begin{proof}
Recall that by \eqref{exact egorov for T(n)}, since 
$n\cdot \iota(\beta)=\iota(\nu\beta)$ for $\beta\in \OF$, $n=\iota(\nu)$,   
$$
\w(\iota(\beta))^{-1} \T(n) \w(\iota(\beta)) = \TN(\iota(\nu\beta))
$$
Also note that $\T(w)^* = \T(-w)$ for all $w$ by \eqref{adjoint T(n)}.  
Substituting the definition of $\D$ and expanding we
  see that $ (\D^{*}\D)^2 $ is given by $1/|\Torusev(2N)|^4$ 
times a sum, ranging over all $\beta_1,
  \beta_2, \beta_3, \beta_4 \in \Torusev(2N)$, of terms
\begin{multline*}
 \TN(\iota(\nu \beta_1))\T(-\iota(\nu \beta_2))\T(\iota(\nu
  \beta_3)) \T(-\iota(\nu \beta_4))   \\=
\gamma(\beta_1,\beta_2,\beta_3, \beta_4)  
\TN( \iota(\nu( \beta_1- \beta_2 +\beta_3- \beta_4)) ) 
\end{multline*}
where $\gamma(\beta_1, \beta_2,\beta_3, \beta_4)$ has absolute
value one (see \eqref{T(m+n)}). Now take the trace; by
Lemma~\ref{l:trace}, the absolute value of the trace of
$\T(n)$ equals $N$ if $\nn \equiv (0,0) \mod N$, zero otherwise. The
result now follows by taking absolute values and summing over all
$\beta_1, \beta_2, \beta_3, \beta_4 \in \Torusev(2N)$.
\end{proof}


%

It remains to estimate the number of solutions of 
\begin{equation}
\label{e:N-congruence}
\nu (\beta_1-\beta_2+\beta_3-\beta_4)  \equiv 0 \mod NI, \quad 
\beta_i \in \Torusev(2N).
\end{equation}
We will show 
\begin{prop}\label{prop-count-sols}
The number of solutions to equation~\eqref{e:N-congruence}  
is bounded by  $O( |\mnorm(\nu)|^{8} N^{2+\epsilon})$. 
\end{prop}

\subsection{Proof of Theorem~\ref{main thm for onb's for T(N)}: Conclusion} 
By Proposition~\ref{prop reduce to count}, we need a suitable upper
bound for the number of solutions of equation \eqref{e:N-congruence},
and a lower bound for the number of elements of $\Torusev(2N)$. 
By Proposition~\ref{prop-count-sols}, the
number of solutions is at most $|\mnorm(\nu)|^8 N^{2+\epsilon}$. 
Note that $|\mnorm(\nu)| \ll |n|^2$. 
  From Lemma~\ref{number of Hecke ops} we 
obtain that $|\Torusev(2N)| \gg N^{1-\epsilon}$ and the result 
follows.

\newpage
\section{Counting solutions}\label{sec: counting}
In this section, we prove Proposition~\ref{prop-count-sols}.

\subsection{A reduction}
\label{s:recover}
%
%
Since $NI \subseteq N\OF\subseteq N\OO$, the number of solutions to
\eqref{e:N-congruence} is bounded by the number of solutions to
$$
\vv (\beta_1-\beta_2+\beta_3-\beta_4)  \in  N \OO, \ \ 
\beta_i \in \Torusev(2N).
$$
Moreover, at the cost of increasing slightly the number of solutions, 
we may omit the parity condition on $\beta_i$ and so replace
$\Torusev(2N)$ by $\Torus(2N)$.

The inclusion $\OF\subset \OK$ induces a map  $\OF/M\OF \to \OK/M\OK$
which has kernel and co-kernel of size at most $[\OK:\OF]$,
independent of $M$. Therefore the induced map 
\begin{multline*}
\Torus(M)=\ker[(\OF/M\OF)^*\to (\Z/M\Z)^*]\\ 
\to \TorusK(M) = \ker[(\OK/M\OK)^*\to (\Z/M\Z)^*]
\end{multline*}
on norm-one elements also has
bounded kernel and co-kernel. 
Thus, up to a bounded factor (depending on $A$ but not on $N$ or
$\nu$), 
%
%
the number of solutions to \eqref{e:N-congruence} is bounded by 
the number of solutions of 
\begin{equation}\label{eqnNormnu}
\nu (\beta_1-\beta_2+\beta_3-\beta_4) =0\mod  N \OK, \quad 
\beta_i \in \TorusK(2N )
\end{equation}

At the cost of increasing the number of solutions, we multiply the
equation \eqref{eqnNormnu} by the Galois conjugate $\bar\nu$ 
to get an equation 
$$
\mnorm(\nu) (\beta_1-\beta_2+\beta_3-\beta_4)  =0\mod  N \OK, \quad 
\beta_i \in \TorusK(2N )
$$
Setting $$N'= \frac N{\gcd(N,\mnorm(\nu))}$$ 
this equation is equivalent to 
\begin{equation}\label{eqnzero}
\beta_1-\beta_2+\beta_3-\beta_4=0  \mod N'\OK, \quad 
\beta_i \in \TorusK(2N)
\end{equation}

Next, note that the reduction map $\OK/rs\OK\to \OK/r\OK$ has kernel
$r\OK/rs\OK \simeq \OK/s\OK$ of size $s^2$, and so the induced map on
norm-one elements $\TorusK(rs)\to\TorusK(r)$ has kernel of order at
most $s^2$ (this is crude, but sufficient for our purposes). 
Thus the reduction map $\TorusK(2N)\to\TorusK(N')$
has kernel of size at most $4\gcd(N,\mnorm(\nu))^2\leq 
4|\mnorm(\nu)|^2$. Therefore the number of solutions of
\eqref{eqnzero} is bounded 
by  $(4|\mnorm(\nu)|^2)^4$ times the number of solutions of the
equation 
\begin{equation}\label{eqnbefone}
\beta_1-\beta_2+\beta_3-\beta_4=0  \mod  N'
\OK, \quad \beta_i \in \TorusK(N')
\end{equation}

 Equation \eqref{eqnbefone} is invariant under Galois
conjugation and we obtain 
a second equation (note that $\bar{\beta}=\beta^{-1}$ since
$\mnorm(\beta)=1\mod N'$) 
\begin{equation}\label{eqnbeftwo}
\beta_1^{-1}-\beta_2^{-1}+\beta_3^{-1}-\beta_4^{-1} 
\equiv 
0 \mod N' \OK.
\end{equation}

\subsection{A transformation} 
We thus have a system of equations \eqref{eqnbefone},
\eqref{eqnbeftwo}, which we transform using the following:
\begin{lem}
\label{l:massage}
If $x,y,z,w$ are invertible then the system of equations
$$
\begin{cases}
x+y=z+w \\
x^{-1}+y^{-1}=z^{-1}+w^{-1} 
\end{cases}
$$
is equivalent to the system 
$$
\begin{cases}
(z-x)(z-y)(x+y) = 0 \\
w=x+y-z
\end{cases}
$$
\end{lem}
\begin{proof}
 From the second equation we get
$$
\frac{x+y}{xy}=\frac{z+w}{zw},
$$ 
or
$$
(x+y)zw=(z+w)xy
$$
The first equation gives that $w=x+y-z$, inserting it in
$(x+y)zw=(z+w)xy$ we get
$$
(x+y)z(x+y-z)=(x+y)xy
$$
or
$$
0=(x+y)(zx+zy-z^2-xy) =-(z-x)(z-y)(x+y).
$$
\end{proof}

Thus by lemma~\ref{l:massage}  the system of equations
\eqref{eqnbefone}, \eqref{eqnbeftwo} is equivalent to
the system: 
\begin{gather}
\label{e:massaged}
(\beta_3-\beta_1)(\beta_3-\beta_2)(\beta_1+\beta_2) \equiv
0 \mod N'\OK, \\
\label{e:massaged-two'}
\beta_4 \equiv \beta_1-\beta_2+\beta_3  \mod N'\OK.
\end{gather}
with $\beta_i\in \TorusK(N')$. 


Since $\beta_4$ is determined by $\beta_1,\beta_2,\beta_3$, we may 
ignore the second equation \eqref{e:massaged-two'} (at the cost of
increasing the number of solutions, since being in $\TorusK(N')$ is a
non-empty condition). 
Multiplying equation~\eqref{e:massaged} by $\beta_3^{-3}$ and letting
$\beta_i' = \beta_i/\beta_3$ we obtain
\begin{gather}
\label{e:massaged-two}
(1-\beta_1')(1-\beta_2')(\beta_1'+\beta_2') \equiv
0 \mod N'\OK
\end{gather}
Since $\beta_3$ is arbitrary, the number  of solutions of
\eqref{e:massaged} is bounded by $|\TorusK(N')|$ times the number of
solutions in $\beta'_1,\beta'_2\in \TorusK(N')$ to \eqref{e:massaged-two}.

\subsection{Prime powers}
By the Chinese remainder theorem the number of solutions to 
\eqref{e:massaged-two} is
multiplicative, and we may concentrate on the prime power case. 
Thus we need to count the solutions to the equation 
\begin{equation}\label{finaleq p^k}
(1-\beta_1')(1-\beta_2')(\beta_1'+\beta_2') \equiv
0 \mod p^k\OK
\end{equation}
with $\beta'_i\in \OK/p^k\OK$, $\mnorm(\beta'_i)=1\mod p^k$. 

We first recall some properties of primes in
quadratic extensions: 
Let $\PP|p$ be a prime in $\OO$ lying above $p$, and let $e$ denote
the ramification index, i.e. the largest integer $e$ such that $\PP^e
| p \OO$. Since $K$ is quadratic $e \in\{1,2\}$, and 
$e=1$ for all but finitely many primes $p$.  
If $e=2$ then $p$ is said to be {\em ramified.} If $e=1$
then $p$ is called unramified, and one of two things can happen:
either $p\OO = \PP$ is still a prime ideal, 
in which case $p$ is said to be {\em inert}, or
$p \OO = \PP \overline{\PP}$, in which case $p$ is said to 
{\em split}.

Now, fix a prime $p$ with ramification index $e$, be it one or two.
The norm map $\mnorm : \OO \rightarrow \Z$ gives a well-defined
homomorphism 
$$
\left( \OO/\PP^{ek} \right)^\times \rightarrow 
\left( \Z/p^{k} \right)^\times.
$$ 
We let
$$
\left( \OO/\PP^{ek} \right)^1
$$ 
be the kernel of this map, i.e., the group of {\em norm one elements}.
For $l \leq ek$ we let
$$
\left( (1+\PP^l)/(1+\PP^{ek}) \right)^1
$$ be the norm one elements in the subgroup $(1+\PP^l)/(1+\PP^{ek})$,
these are precisely the norm one elements that reduce to one modulo
$\PP^l$.

\begin{lem}\label{count sols of finaleq p^k}
There is a constant $c>1$ so that the number of solutions of
equation~\eqref{finaleq p^k} is at most $ck p^k$. 
\end{lem}

\begin{proof}
Equation \eqref{finaleq p^k} is invariant under
Galois conjugation, therefore its solutions in $\OO/p^k\OO$ correspond
bijectively to solutions  $\beta'_i\in \OO/\PP^{ek}$, $\mnorm(\beta'_i)=1
\mod p^k$ (this is of course only an
issue in the split case where $\OO/p^{k}\OO \simeq \OO/\PP^{k} \times
\OO/\overline{\PP}^{k}$). Thus we need to count solutions of 
\begin{equation}\label{eqnPP^ke}
(1-\beta_1')(1-\beta_2')(\beta_1'+\beta_2') \equiv
0 \mod \PP^{ek} 
\end{equation}
with $\beta'_i \in \OO/\PP^{ek}$, $\mnorm(\beta'_i)=1\mod p^k$.  

We will first assume that $p$ is {\em odd}. Since $\beta_1' \equiv
\beta_2' \equiv 1 \mod \PP$ implies that $\beta_1'+\beta_2' \equiv 2
\not \equiv 0 \mod \PP$ we see that at most two of the factors in
equation~\eqref{eqnPP^ke} can be congruent to zero modulo
$\PP$. Moreover, we may assume that the third factor is nonzero by
multiplying by a 
suitable $\beta$ and permuting the variables. (Of course we must then
compensate by multiplying the number of solutions by $\binom{3}{2}$). 
Now, if the product is zero modulo $\PP^{ek}$, then there is some $0
\leq n \leq ek$ such that one factor is zero modulo $\PP^n$, and the
other zero modulo $\PP^{ek-n}$. 
Thus 
the number of solutions to equation~\eqref{eqnPP^ke} equals
$$
\binom{3}{2} 
\sum_{n=1}^{ek-1}
\left| \left( (1+\PP^n)/(1+\PP^{ek}) \right)^1  \right|
\times  
\left| \left( (1+\PP^{ek-n})/(1+\PP^{ek}) \right)^1  \right| +
$$
$$
2 \left| \left( \OO/\PP^{ek} \right)^1 \right|
$$
Using Lemma~\ref{l:norm-one-reduced} we obtain
$$
\left| \left( (1+\PP^n)/(1+\PP^{ek}) \right)^1  \right|
\times  
\left| \left( (1+\PP^{ek-n})/(1+\PP^{ek}) \right)^1  \right|
\leq 
p^{k+e-1}
$$
and by Lemma~\ref{l:number-of-norm-one}
$$
\left| \left( \OO/\PP^{ek} \right)^1 \right|
\leq 
2(p+1)p^{k-1}.
$$
Hence for $p$ odd, the total number of solutions to \eqref{eqnPP^ke}  
is bounded by
$$
4(p+1)p^{k-1} + 3 (ek-1) p^{e-1} p^k \ll kp^k
$$ 
(since $e=1$ for all but finitely many primes). 

If $p=2$ it is no longer true that only two factors can be
zero modulo $\PP$. However, $\beta_1 \equiv \beta_2 \equiv 1 \mod
\PP^{e+1}$ implies that $\beta_1+\beta_2 \equiv 2 \mod \PP^{e+1}$.
Since $2\OO=\PP^e$, we see that if two factors are zero modulo
$\PP^{e+1}$, then the third factor can be congruent to zero at most
modulo $\PP^{e}$. We may thus bound the number of solutions by
counting the number of ways the product of two factors can be equal to
zero modulo $ \PP^{ek-e}$. This we can do as we did for odd primes,
and we obtain the same bound as before, except that we lose an
additional factor of at most
$$
\left| \left( (1+\PP^{ek-e})/(1+\PP^{ek}) \right)^1 \right|^4
\ll
2^{O(e)} = O(1).
$$
This proves the Lemma.
\end{proof}

%

\subsection{Proof of Proposition~\ref{prop-count-sols}} 
By multiplying over all primes, we  see from 
Lemma~\ref{count sols of finaleq p^k} 
that the number of solutions of equation~\eqref{e:massaged-two} is   
$O( (N')^{1+\epsilon})$. Therefore we see that the number of
solutions of \eqref{e:massaged} is  
$O((N')^{2+\epsilon})$ since $|\TorusK(N')|\ll (N')^{1+\epsilon}$ by
Lemma~\ref{l:number-of-norm-one}. 
This gives a bound for the solutions of \eqref{eqnbefone} and
multiplying by $|\mnorm(\nu)|^8$ gives a bound for the number of
solutions of \eqref{eqnzero}. In turn, by the reasoning in
section~\ref{s:recover}  this gives 
a bound of $O(|\mnorm(\nu)|^8 N^{2+\epsilon})$ on the solutions of
\eqref{e:N-congruence}.






\newpage
\appendix
\section{An identity of Gauss sums} \label{appgauss}

For section~\ref{multiplicativity} we need to prove the identity 
\eqref{weird gauss sum identity}. 
To prove it we will need a lemma about Gauss sums. Given an integer
$x$, we define its ``dyadic valuation'', $v(x)$, by $x = 2^{v(x)}
x_0$, where $x_0$ is an odd integer. Let 
$$
G(b,c) =  \sum_{ z \mod 2^{k}} e \left( \frac{\rr}{2^{k}}( -bz^2 +
  2cz) \right).
$$ 

\begin{lem}
\label{l:gauss-sum}
If $ v(c) < v(b) < k$ then 
$$
G(b,c) = \begin{cases}
2^k & \text{if $v(b)=k-1$ and $v(c)=k-2$,} \\
0 & \text{otherwise.}
\end{cases}
$$
\end{lem}
\begin{proof}
We may write 
$$
G(b,c) = 
\sum_{ z \mod 2^{k}} e \left( \frac{2c \rr} {2^{k}}( -\beta z^2 + z) \right)
$$ 
where $\beta$ is an integer satisfying $2c \beta \equiv b \mod 2^{k}$.
Let $n=k-1-v(c)$;
it is the smallest integer $n$ such that $e \left( \frac{2 c \rr
    }{2^{k}} x \right) = 1$ for all $x \equiv 0 \mod 2^n$.

Assume first that $n>1$. Let $\epsilon = \epsilon_0 2^{n-1}$ be such
that $e \left( \frac{2c \rr }{2^{k}} \epsilon \right) \neq 1$. 
Making the change of variables $z \rightarrow z+\epsilon$ we see that
\begin{equation*}
\begin{split}
G(b,c) &=
\sum_{ z \mod 2^{k}} 
e \left( \frac{2c \rr}{2^{k}}( -\beta ( z^2 + 2 \epsilon z + \epsilon^2)
+ z+\epsilon) \right)\\
&= G(b,c) e \left( \frac{2c \rr }{2^{k}} \epsilon \right)
\end{split}
\end{equation*}
since $2 \epsilon z + \epsilon^2 \equiv 0 \mod 2^n$. But
$e \left( \frac{2c \rr }{2^{k}} \epsilon \right) \neq 1$ and
therefore $G(b,c) = 0$.  

If $n \leq 1$ then, as $n=k-1-v(c)$ and $v(c)<v(b) < k$, we must have
$n=1, v(c)=k-2$ and $v(b)=k-1$. Hence $\beta \equiv 1 \mod 2$.
Moreover, if $n=1$ we must have $e \left( \frac{2c \rr x}{2^{k}}
\right) =  e \left( \frac{x}{2} \right)$. 
Thus
$$
G(b,c) =
\sum_{ z \mod 2^{k}} 
e \left(\frac{z^2+z}{2} \right) =
2^k
$$
since $z^2+z \equiv 0 \mod 2$ for all $z$. 
\end{proof}

\begin{prop} The following equality holds for all $x,y$
\begin{multline*}
\sum_{ z \mod 2^{k}} 
e \left( \frac{\rr}{2^{k}} (- bz^2 + 2z(y-dx) + a d^2 x^2) \right)\\
=
\sum_{ z \mod 2^{k}} 
e \left( \frac{\rr}{2^{k}} (-d^{-1}bz^2 + 2z(y-x) + a d y^2) \right).
\end{multline*}

\end{prop}

\begin{proof}
  The case $v(b) \geq k$, i.e. $b \equiv 0 \mod 2^k$, implies that $d
  \equiv 1 \mod 2^k$ and equality holds trivially. We may thus assume
  that $v(b) < k$. 

  We begin by noting that since $ y-dx = d( d^{-1}y - x) = d(y-x+aby) $
  we see that $v(y-x) < v(b)$ implies that $v(y-dx) < v(b)$; putting
  $x'=d^{-1}x$ we see that the converse holds, and hence $v(y-x) <
  v(b)$ if and only if $v(y-dx) < v(b)$.

  {\bf First case, $v(y-x) < v(b)$:} 
  Putting $c=y-x$, $c=y-dx$ respectively and applying
  lemma~\ref{l:gauss-sum} we see that both sides are zero except 
  when $v(c)=k-2$ and $v(b)=k-1$. For the exceptional case we note
  that $v(b)=k-1$ implies that $d^{-1} = 1+ab \equiv 1 \mod 2^{k}$,
  and the same holds for $d$. Moreover, $v(c) = k-2$ means that $x
  \equiv y \mod 2^{k-2}$ and since $4|a$ we see that

  $$ LHS = 2^k e \left( \frac{\rr}{2^{k}} ad^2x^2 \right) = 2^k e
  \left( \frac{\rr}{2^{k}} ady^2 \right) = RHS.
  $$

  {\bf Second case, $v(y-x) \geq v(b)$:} As remarked above this means
  that $v(y-dx) \geq v(b)$. We may thus complete the squares inside
  the exponentials, and we get
  $$ LHS = \sum_{ z \mod 2^{k}} e \left( \frac{\rr}{2^{k}} \left( - b(
      z - \frac{y-dx}{b} )^2 + \frac{(y-dx)^2}{b} + a d^2 x^2
    \right)\right)
  $$ 

  and

  $$ RHS = \sum_{ z \mod 2^{k}} e \left( \frac{\rr}{2^{k}} \left( -
      d^{-1}b( z - \frac{d(y-x)}{b} )^2 + \frac{d(y-x)^2}{b} + a d y^2
    \right) \right).
  $$

  After changing variables and taking constants outside we get

  $$ LHS = e \left( \frac{\rr}{2^{k}} \left( \frac{(y-dx)^2}{b} + a
      d^2 x^2 \right) \right) \sum_{ z \mod 2^{k}} e \left(
    \frac{\rr}{2^{k}} \left( - b z^2 \right) \right)
  $$ 

  and
  $$ RHS = e \left( \frac{\rr}{2^{k}} \left( \frac{d(y-x)^2}{b} + a d
      y^2 \right) \right) \sum_{ z \mod 2^{k}} e \left(
    \frac{\rr}{2^{k}} \left( - d^{-1}b z^2 \right) \right).
  $$

  Now, $d \equiv 1 \mod 16$ means that $d$ is a square modulo
  $2^k$. Changing variables by $z \rightarrow \sqrt{d}z$ in the second
  sum we see that the sums are equal, and we are left to prove that
$$
e \left( \frac{\rr}{2^{k}} \left( 
  \frac{(y-dx)^2}{b} + a d^2 x^2
\right) \right)
=
e \left( \frac{\rr}{2^{k}} \left( 
  \frac{d(y-x)^2}{b} + a d y^2
\right) \right).
$$
This will follow from the equality
$$
\frac{(y-dx)^2}{b} + a d^2 x^2 
=
\frac{d(y-x)^2}{b} + a d y^2.
$$
Collecting terms it is equivalent to 
\begin{equation*}
\begin{split}
0 &= ad ( y^2 - d x^2 ) + 
b^{-1}( dy^2 + dx^2 - 2dxy - y^2 - d^2 x^2 + 2dxy )\\
&= 
ad ( y^2 - d x^2 ) + 
b^{-1}( y^2(d-1) + x^2( d-d^2 ) )\\ 
&=
ad ( y^2 - d x^2 ) + 
(d-1)b^{-1}( y^2 - d x^2 ),
\end{split}
\end{equation*}
which follows from the identity
\begin{equation*}
\begin{split}
ad + (d-1)/b &=
d( a + \frac{1-1/d}{b} )\\
& =d( a + \frac{1-(1+ab)}{b} )\\
& =d( a - \frac{ab}{b} ) = 0.
\end{split}
\end{equation*}
\end{proof}


\newpage

\section{Counting norm one elements}\label{app: count}

Let $e$ be the ramification index of a prime $p$ in $\OO$, i.e. the
largest integer such that $\PP^e | p \OO$, where $\PP \subset \OO$ is
any prime ideal dividing $p \OO$.  Since $K$ is quadratic $e \in
\{1,2\}$. If $e=2$ then $p$ is said to be {\em ramified.} If $e=1$
then $p$ is called unramified, and one of two things can happen:
either $p\OO = \PP$, in which case $p$ is said to be {\em inert,} or
$p \OO = \PP \overline{\PP}$, in which case $p$ is said to {\em
  split}.

Now, fix a prime $p$ with ramification index $e$, be it one or two.
The norm map
$$
\mnorm : \OO \rightarrow \Z
$$ 
descends modulo $p^k$, and gives a homomorphism
$$
\left( \OO/\PP^{ek} \right)^\times
\rightarrow
\left( \Z/p^{k} \right)^\times.
$$ 
%
We let
$$
\left( \OO/\PP^{ek} \right)^1
$$ 
be the kernel of this map, i.e., the group of {\em norm one elements}.
For $l \leq ek$ we let
$$
\left( (1+\PP^l)/(1+\PP^{ek}) \right)^1
$$ be the norm one elements in the subgroup $(1+\PP^l)/(1+\PP^{ek})$,
these are precisely the norm one elements that reduce to one modulo
$\PP^l$.

\begin{lem}
\label{l:number-of-norm-one}
We have
$$
\left| \left( \OO/\PP^{ek} \right)^1 \right|=
\begin{cases}
(p-1)p^{k-1} & \text{if $p$ is split,} \\
(p+1)p^{k-1} & \text{if $p$ is inert,} \\
2p^{k} & \text{if $p$ is ramified.} 
\end{cases}
$$
\end{lem}
\begin{proof}
Recall first from class field theory \cite{CF} that the index (in
$\Z_p^\times$) of the image of the units in the $p$-adic completion
of $\OO$ under the norm map equals the ramification index $e$. 
We will split the proof in three parts: 

{\em The split case:} 
If $p$ splits in $K$ then $p \OO = P_1 P_2$ where $P_1,P_2$ are prime
ideals in $\OO$, and where $P_2 = \overline{P_1}$.  The map $x
\rightarrow \overline{x}$ gives an isomorphism between $\OO/P_1^k$ and
$\OO/P_2^k$. This, together with the Chinese Remainder Theorem gives 
$$
\OO/p^k\OO 
\simeq
\OO/P_1^k \times \OO/P_2^k 
\simeq
\OO/P_1^k \times \OO/P_1^k 
$$ where $x \in \OO/p^k\OO$ is mapped to $(x,\overline{x}) \in
\OO/P_1^k \times \OO/P_1^k$. Furthermore, $\OO/P_1^k \simeq \Z/p^k\Z$,
and therefore 
\begin{equation}
\label{e:split-isomorphism}
\OO/p^k\OO 
\simeq
\Z/p^k\Z \times \Z/p^k\Z.
\end{equation}
Under this isomorphism, Galois conjugation maps $(x,y) \in
\Z/p^k\Z \times \Z/p^k\Z$ to $(y,x)$. Thus the natural embedding of
$\Z/p^k\Z$ in $\OO/p^k\OO \simeq \Z/p^k\Z \times \Z/p^k\Z$ consists of
elements of the form $(x,x)$, and the image of $(x,y)$ under the norm
map is $(xy,xy)$. Hence the norm one elements in $\OO/p^k\OO$
correspond to elements of the form $(x,y) \in \Z/p^k\Z \times
\Z/p^k\Z$ such that $xy=1$, and the number of such elements is
$(p-1)p^{k-1}$.

{\em The inert case:} 
%
%
Here $e=1$ and the local norm map is onto $\Z_p^\times$; reducing
modulo $p$ we get an exact sequence
$$
1 
\rightarrow
(\OO/\PP^{k})^1
\rightarrow
(\OO/\PP^{k})^\times
\rightarrow
(\Z/p^k)^\times
\rightarrow
1.
$$
Hence
$$
|(\OO/\PP^{k})^1| = 
\frac{|(\OO/\PP^{k})^\times|}{|(\Z/p^k)^\times|}
= (p+1)p^{k-1}.
$$

{\em The ramified case:} Here the image of the norm map in
$\Z_p^\times$ is of index 2, and thus the image of the norm in
$(\Z/p^k)^\times$ has cardinality $\frac{(p-1)p^{k-1}}{2}$.
Consequently,
$$
|(\OO/\PP^{ek})^1| = 
2 \frac{ |(\OO/\PP^{ek})^\times| }{ (p-1)p^{k-1} }.
$$
Now,
$$
| (\OO/\PP^{ek})^\times |
=
| (\OO/\PP)^\times | \times
| (1+\PP)/(1+\PP^{ek}) |
=
(p-1) p^{ek-1}
$$
and since $e=2$ we get
$$
|(\OO/\PP^{ek})^1| = 
2 \frac{ (p-1) p^{2k-1} }{ (p-1)p^{k-1} } =
2 p^{k}.
$$
\end{proof}

We will also need to know the number of norm one elements that reduce
to one modulo $\PP^l$. 
\begin{lem}
\label{l:norm-one-reduced}
We have
$$
\left| \left( (1+\PP^l)/(1+\PP^{ek}) \right)^1  \right|=
\begin{cases}
  p^{ k-l } & \text{if $p$ is split or inert,} \\ 
  K_p \times p^{ k+\lceil l/2 \rceil - l } & \text{if $p$ is ramified.}
\end{cases}
$$
where $K_p = 1$ if $p$ is odd, and $K_2 = 1$ or $2$.
\end{lem}
\begin{proof}

{\em The split case:} 
 From the previous discussion of the isomorphism in
equation~\ref{e:split-isomorphism} we see that norm one elements
congruent to one modulo $P_1^l$ correspond to elements $(x,x^{-1}) \in
\Z/p^k\Z \times \Z/p^k\Z$ such that $x \equiv 1 \mod p^l$. The number
of such elements is $|(1+p^l)/(1+p^k)|=p^{k-l}$.

{\em The inert case:} If $p$ is {\em odd} then $x\rightarrow x^2$ is
an automorphism of 
$(1+\PP^l)/(1+\PP^k)$ since the order of the group is odd. Thus the norm
is locally onto in the sense that the map
$$
\mnorm :
(1+\PP^l)/(1+\PP^k) 
\rightarrow 
(1+p^l)/(1+p^k) 
$$
is onto.

If $p$ is {\em even} (and inert) then squaring is not an automorphism as
$(1+x)^2= 1+2x+x^2$. However, $1+p^l \subset 1+\PP^l$ and 
squaring maps $(1+p^l)/(1+p^k)$ onto $(1+p^{l+1})/(1+p^k)$. Thus
$$
(1+p^{l+1})/(1+p^k) 
\subset 
\mnorm \left((1+\PP^l)/(1+\PP^k) \right),
$$
which shows that the image of the norms must be either
$(1+p^{l+1})/(1+p^k)$ or $(1+p^{l})/(1+p^k)$. (There are no subgroups
in between!) We will show that the former holds; since $2$ is
unramified the discriminant of $K$ is odd and $\OO
=\Z[\frac{1+\sqrt{d_k}}{2}]$. Hence $\trace(\OO) = \Z$, and there
exists $x \in \OO$ with odd trace. Now,
$$
\mnorm( 1+p^k x) = 1 + p^k \trace(x) + p^{2k} \mnorm(x),
$$ 
shows that the image must be $(1+p^{l})/(1+p^k)$.

Thus, whether $p$ is even or odd, the norm map is locally onto and
hence
%
$$
\left| \left( (1+\PP^l)/(1+\PP^k) \right)^1 \right| = 
\frac{ \left| \left( (1+\PP^l)/(1+\PP^k) \right)^\times \right| }
{\left| \left( (1+p^l)/(1+p^k) \right)^\times  \right|}
= p^{k-l}.
$$

{\em The ramified case:} First we note that 
\begin{equation}
\label{e:inclusion}
\mnorm \left( (1+\PP^l)/(1+\PP^{ek}) \right) \subset
(1+p^{\lceil l/2 \rceil})/(1+p^{k}).  
\end{equation}

Arguing as before that squares are in the image of the norm we see
that equality holds for {\em odd} $p$, and we obtain
$$
\left| \left( (1+\PP^l)/(1+\PP^{ek}) \right)^1 \right| = 
\frac{ \left| \left( (1+\PP^l)/(1+\PP^{ek}) \right)^\times \right| }
{\left| \left( (1+p^{\lceil l/2 \rceil})/(1+p^{k}) \right)^\times  \right|}
= 
$$
$$
\frac{ \left| \OO/\PP \right|^{2k-l}    }
{ p^{k-\lceil l/2 \rceil}} 
=
\frac{ p^{2k-l} } { p^{k-\lceil l/2 \rceil} }
=
p^{k+\lceil l/2 \rceil -l}
$$

For $p$ {\em even} the squaring argument shows that 
$$
(1+p^{\lceil l/2 \rceil +1} )/(1+p^{k})
\subset
\mnorm \left( (1+\PP^l)/(1+\PP^{ek}) \right),
$$ 
which gives a lower bound on the image. This gives the same result as
for the odd case, except for a factor of $2$. 
\end{proof}



\newpage
\bibliographystyle{amsplain}

\end{document}